# Hole/Electron transport layers in tin-doped SBLN *nano* materials for hybrid solar cells

**Anurag Pritam and Vaibhav Shrivastava**

Dielectric Laboratory, Shiv Nadar University, G.B.Nagar-201314, Uttar Pradesh, India

*In this work, layered perovskite SBN was investigated in a new doped form for hole as well as electron transport layer (HTL/ETL) in perovskite solar cells. This work was targeted to conclude the effect of tin doping in lanthanum-bismuth layer SBN on optical energy band gap besides dominant electron-hole transportation to assist in perovskite solar cell applications. Thoroughly hard ball-milled compositions $Sr_{1-x}Sn_xBi_{1.95}La_{0.05}Nb_2O_9$ (x=0.0, 0.01, 0.03, 0.05, 0.1 and 0.2) were prepared by special microwave synthesis to obtain fine (~10-60nm) mesoporous particle network of atomic level substitutions. Microwave synthesis was crucial in modifying dielectric, semiconducting and optical characteristics of prepared SBN materials. The band gap reduced in continuous manner and carrier mobility was increased by 112% for maximum tin doping. Nano particle formation assisted in raising carrier mobility by bridging bigger grains through nano particles. The effect of macro-sized grains and nano-sized grain boundaries on carrier transport were further investigated in detail using impedance spectroscopy.*

## Introduction

In recent years, perovskites ($ABX_3$; X-oxides and halides) have emerged as multitasking materials not only limited to conventional ferroelectric and piezoelectric applications but also like promising pyroelectric and solar cell's hole transport medium (HTM) [1-6] as well. Many researchers have highlighted the successful utilization of *lattice strain sensitive* perovskite materials into solar cells with high power conversion efficiency (PCE) ~ 22.1% [5-7]. O'Regan and Grätzel [7] have developed a prototype of low cost perovskite solar cells by modifying large surface area of nanocrystalline $TiO_2$ films through dye. This *photo electrochemical* version of solar cell prototype suffer with problems like charge leakage in liquid electrolyte. They propose the solution by using *nano crystalline* TiO$_2$ films containing pores of sizes 5-50*nm* and capacity to act as hole transport layer (HTL). There is no difference between liquid electrolyte based dye-sensitized solar cells and TiO$_2$ based dye-sensitized solar cells other than interrupted direct contact between transparent conducting oxide (TCO) layer and hole transporting medium (HTM). The spiro-MeOTAD is preferred for drop casting on nanocrystalline TiO$_2$ layer enabling heterojunction formation to generate photocurrent. Current work is targeted to prepare *mesoporous* bismuth-lanthanum layered perovskite Sr$_{1-x}$Sn$_x$Bi$_{1.95}$La$_{0.05}$Nb$_2$O$_9$ (SSBLN) semiconducting layers working like HTL *as well as* ETL in conjugation with suitable dye for offering remarkably high PCE. These bismuth layered Aurivillius materials have attracted researchers [8], structural distortions sensitive dielectric, ferroelectric and spectroscopic response driving multitasking in these materials. The vulnerability of bismuth oxide layer at high working temperatures above 825°C is the main cause of restricted use of Aurivillius materials. The loss of bismuth oxide releases *c*-axis strain and renders almost no control of bismuth oxide layer over perovskite octahedrons. Researchers [9] have attempted various trials like hydrogen bonding in perovskite halides, using toxic lead as dopant in perovskite oxides to increase modulus of electric polarization at the cost of unaltered *c*-axis lattice strain. We target in present work to prepare a novel tin-doped lanthanum controlled bismuth layer SrBi$_2$Nb$_2$O$_9$ (SBN). The proposed composition is Sr$_{1-x}$Sn$_x$Bi$_{1.95}$La$_{0.05}$Nb$_2$O$_9$ (SSBLN) where after number of trials doping lanthanum onto bismuth sites by 5% was found optimum for checking mixture weight loss up to 1000°C due to bismuth oxide evaporation. In present work, *p*-orbital Sn$^{2+}$ is substituted in place of *s*-orbital Sr$^{2+}$ to generate more polarity in the crystal system besides converting conventionally known insulating behavior into semiconducting for potential photovoltaic applications. The prepared mesoporous SSBLN compositions are characterized for smooth migration of charges among grains without being perturbed by capacitive grain boundaries and electrode-material interface along with expected increased in optical absorption and Hall mobility for perovskite solar cell applications.

## Experimental procedure

Reagent-grade oxide and carbonate powders of $SrCO_3$, $SnO$, $Bi_2O_3$, $La_2O_3$ and $Nb_2O_5$ (all from Sigma-Aldrich with purity > 99.99 %) were used as the starting materials to prepare desired $Sr_{1-x}Sn_xBi_{1.95}La_{0.05}Nb_2O_9$ (SSBLN) ceramic compositions. The stoichiometric proportions of these powders were taken for dry milling up to 6h, 12h and 20h in a planetary ball mill Retsch PM200 using *hard* tungsten carbide balls of diameter 5mm (the ball to powder mass ratio was 8:1). The obtained powders were admixed with 4.0wt% polyvinylalcohol (PVA) binder solution and were pressed into circular disks of 10.8 mm diameter using uniaxial pressure of 120MPa. The obtained pellets were single step microwave sintered at 1000°C for 2 h with a heating rate of 3°C/min. and cooling rate of 2°C/min to minimize density gradients. For this purpose, a microwave furnace operating at 2.45GHz frequency based on a pair of magnetrons consuming power 2.2KW was used. The crystal structure of each sample was investigated by X-ray diffraction (XRD) using $CuK_\alpha$ radiation (D8 Advance, Bruker, Germany), in the $2\theta$ range of 10-80° (step increment-0.02° with a time duration of 2sec/step). Nano particle size formation in ball-milled SSBLN compositions was investigated using competent transmission electron microscopy (TEM). For this purpose, typical ($x$=0.0 and 0.2) doped SSBLN composition pellets were crushed and converted powders were used for investigating crystallinity and particle size. TEM images were taken using the Carl Zeiss make LIBRA 200FE high resolution TEM with information limit of 0.13nm equipped with EDS andit operated at 200 kV. Optical energy band gap for each modified SSBLN composition was calculated using Tauc plots. These plots were derived using UV-Vis diffuse absorption spectra measured in an integral sphere mode using spectrophotometer SHIMADZU UV-Vis 3700. For reflecting incident radiations $BaSO_4$ was used as a medium co-mixed with each modified SBN composition. Oxidation-reduction analysis was carried using cyclic voltammetry (CV) performed on Autolab Potentiostat Galvanostat PGSTAT302N (Metrohm, Netherlands). For this electrochemical characterization, CV measurements were recorded in a three-electrode set-up consisting of Ag/AgCl as the reference electrode, platinum wire as the counter electrode and glassy carbon electrode (GCE) modified with individual SSBLN composition as the working electrode. The standard electrolyte used in reaction was 5 mM of Potassium Ferro/ Ferry cyanide in 0.1 M KCl. The CV scans were recorded from -0.4 V to 1V with the scan rate of $0.50 mVs^{-1}$. The vibrational characteristics of all present bonds in investigated SSBLN compositions were studied using Fourier transform infrared (FTIR) spectroscopy. These FTIR spectra were recorded on FTIR Spectrophotometer of Thermo Fisher Scientific make and model Nicolet iS5 in the range of 400 and 4500 $cm^{-1}$. The dielectric properties were measured by using high frequency LCR meter ZM2376 (NF corporation, Japan ) with an applied oscillation level voltage of 1V over the frequency range 20Hz–2MHz. Randomly chosen SSBLN composition thin pellets were used for investing carrier mobility in SSBLN compositions using Hall Effect based on van der Pauw (vdP) method.

**Results and discussion**

All SSBLN composition powders milled up to 20 hours are uniaxially pressed at 120MPa into circular discs and sintered using microwaves at 1000ºC for 2hrs. Perfect perovskite phase is observed in all SSBLN samples, Fig.1 with no secondary phase formation and unreacted phases. All samples show weight loss in the range 0.01%-0.04%, that is significantly lower compared to earlier reported values in these Aurivillius materials due to incorporation of lanthanum onto bismuth sites. These samples have shown weight loss before adding lanthanum in the range 0.1% to 2%. In addition, rigorous hard milling of SBN has helped in avoiding calcination stage and preserve bismuth oxide as much as can be achieved [2-3]. All the diffraction peaks are indexed using orthorhombic symmetry ($A2_1am$) according to the JCPDS card no 00-049-0607. The calculated values of orthorhombic unit cell based lattice parameters were around 5.5073 Å, 5.5060 Å, and 25.0127 Å. The values of lattice parameters *a* and *b* for tin-doped samples are higher by 0.036-0.2% with those of undoped SSBLN whereas, a regularly increasing trend is observed in parameter *c* for all samples leaving the one with 1% tin doping. The values of tetragonal strain (c/a) in the unit cell also provides similar indication, Table-1. Overall, all three lattice parameters show higher values for tin-doped samples compared to sample without tin.

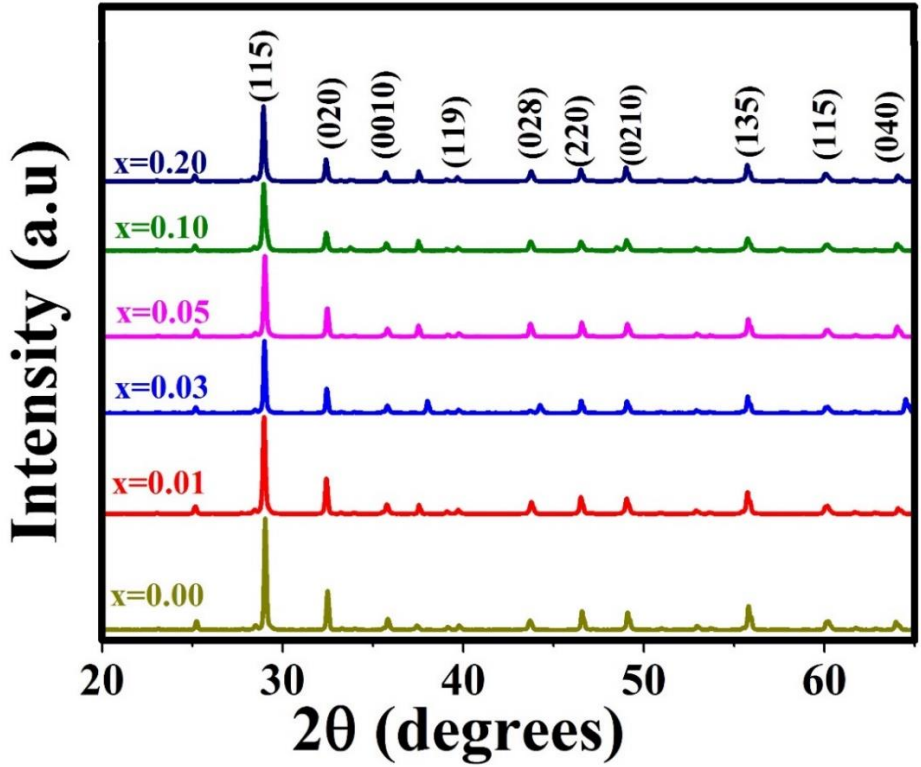

**Fig. 1. XRD patterns of tin-doped SSBLN ceramics microwave sintered at 1000°C for 2 hrs.**

**Table-1 Elements in most stable valence states with ionic radius (IR), coordination number (CN) and bond energy with oxygen. Lattice parameters with variation of tin doping in $Sr_{1-x}Sn_xBi_{1.95}La_{0.05}Nb_2O_9$ ceramics**.

| Elements | IR(Å) | CN | Bond Energy (KJ mol$^{-1}$) | Sample | Unit cell volume (Å$^3$) | c/a strain |
|---|---|---|---|---|---|---|
| $Sr^{2+}$ | 1.44 | 12 | 426 | x = 0.00 | 758.465 | 4.541 |
| $Sn^{2+}$ | 0.96 | 6 | 528 | x = 0.01 | 762.975 | 4.551 |
| $La^{3+}$ | 1.03 | 6 | 798 | x = 0.03 | 760.869 | 4.544 |
| $Bi^{3+}$ | 1.03 | 6 | 336 | x = 0.05 | 762.319 | 4.557 |
| $Nb^{5+}$ | 0.64 | 6 | 703 | x = 0.10 | 763.384 | 4.558 |
|  |  |  |  | x = 0.20 | 764.850 | 4.562 |

The occupancy of tin ($Sn^{2+}$) is critically investigated on various available sites in SSBLN structure based on ionic radius and coordination number as listed in Table-1. An initial (for 3% tin doped sample) *decrease* else overall *increase* in lattice parameters (also unit cell volume) suggest following three probable substitutions: *a*) 1% doping of smaller $Sn^{2+}$ occurs equally on scheduled vacant sites of $Sr^{2+}$ without contributing excess charge and on vacant $Bi^{3+}$ sites (of $Bi_2O_2$ layer), *b*) higher doping (*x* = 3 & 5%) of *p-orbital* $Sn^{2+}$ mostly on vacant sites of $Sr^{2+}$ *imparting* large c-axis displacement through hybridized oxygen's along with 1% $Sn^{2+}$ ions occupying $Bi^{3+}$ sites in bismuth layer and *c*) double $Sn^{2+}$ occupancy on scheduled $Sr^{2+}$ sites *after* saturating all vacant $Bi^{3+}$ sites (beyond 5%) to engage 12 coordination number of site-A of $Sr^{2+}$. The substitution on vacant $Bi^{3+}$-sites should generate more holes to neutralize site valance ($Sn^{2+} + 1h \rightarrow Bi^{3+}$) whereas double tin occupancy on $Sr^{2+}$ sites should generate more electrons ($Sn^{2+} + \underline{Sn^{2+} + 2e^-} \rightarrow Sr^{2+}$). At room temperature, stoichiometric mixing in thoroughly ball milled doped compositions should control site occupancy based on site energy and valence thus $Sn^{2+}$ occupies scheduled $Sr^{2+}$ sites besides randomly distributed in cluster form. X-ray diffractograms confirm the formation of perovskite phase on increasing milling time. As microwave sintering is carried for these milled powders in *compact* pellet form, local atomic level redistribution of $Sn^{2+}$ ions occurs. The formation of perfect perovskite phase, Fig.1, further confirms this one-time thermally activated redistribution of $Sn^{2+}$ ions in three possibilities discussed before in this section. For these possible substitutions, the p-orbital $Sn^{2+}$ occupying s-orbital $Sr^{2+}$ sites along with $Bi^{3+}$ sites (as expected) should result in an increase in *c*-axis strain (c/a) [10]. The same is observed through the values of *c*-axis strain as indicated in Table-1. Such an increase is due to hybridization between Sn 5p states and O 2p orbitals resulting in large oxygen displacement [11] in along *c*-axis. Almost regular increase in unit-cell volume indicates about expected *c*-axis upthrust in octahedral cages of perovskite unit cell. Moreover, it supports possibility of double $Sn^{2+}$ occupancy on $Sr^{2+}$ sites.

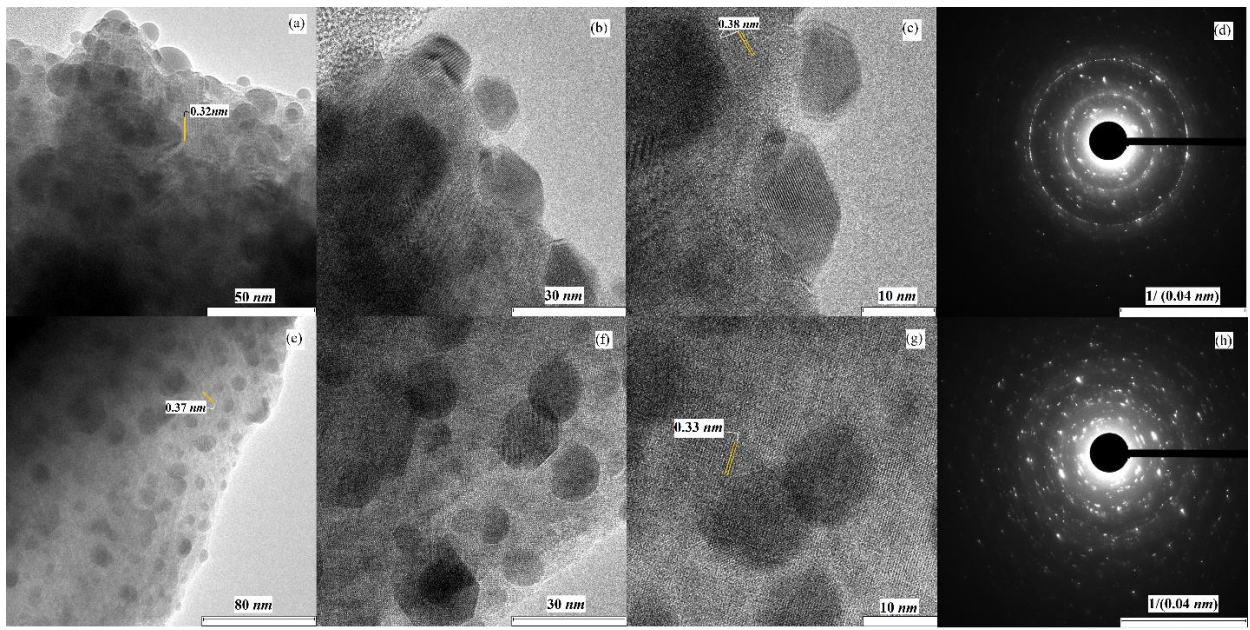

**Fig. 2 TEM images of crushed sintered pellet powders (x = 0.00 and 0.20)**

For checking, the microstructural details, a thorough TEM analysis have been performed on pure and 20% tin-doped SSBLN materials. Figs. 2(a) to (c) represent a typical bright field TEM image of SBLN material without tin indicating formation of small nano particles on bigger grains and shows Moiré fringes [12]. Careful analysis of high magnification image, Fig. 2(b), further confirms that small particles jutted out from the main SSBLN grains due to expected local heating effects of microwave. Now Fig.2(c) represents another magnified portion of tin-less SSBLN particles confirming jutting out of small SSBLN particles from big grains and no co-existence of small and big particles as elusively seen in Fig.2(a). Selected area electron diffraction (SAED) image, Fig.2(d), indicates the presence of both small as well as the bigger grains mixed randomly. The effect of 20 atomic% tin doping is shown in Fig.2(e) confirming that tin addition promoted the formation of small agglomerated grains anchored with bigger grains as in Fig.2(a). Moreover, High magnification images Fig.2 (f) & (g) indicate decrease in *separation* between Moiré pattern formation as seen in case of SSBLN sample without tin [Fig. 2(b) & (c)] confirming the presence of double tin doped SSBLN particle network. Corresponding SAED image, Fig. 2(h), represents an increase in crystallinity of polycrystalline SSBLN materials after doping tin and diffused diffraction spots indicating assumed double tin occupancy. Average grain size was observed to be around 100-150nm in all SSBLN compositions. In addition, average crystallite size (~ 48-62*nm*) estimated from Scherrer's formula is observed to follow good accordance with average particle size 35-48*nm* as estimated from TEM images, Figs.2(a) & (c). The formation of jutted out spherical SSBLN *nano* particles is because of temperature dependent increase in grain size popularly known as Ostwald ripening process [13]. The distribution of energy dispersive spectrometry (EDS) peaks (not shown here) confirm presence of very close atomic stoichiometry as planned with no other elements present.

Moret *et al* [14], reported excellent discussion on infrared activity of twin $SrBi_2Ta_2O_9$ (SBT) material and mentioned about irreducible phonon symmetries delivering 81 modes (A1, A2, B1 and B2). An overall decomposition of modes is, $\Gamma = 21A_1 + 20A_2 + 19B_1 + 21B_2$, where $A_1, B_1$ and $B_2$ are 61 IR-active modes for $E\|a$, $E\|c$ and $E\|b$ respectively. In case of tetragonal SBT, the IR-active modes reduce to thirteen. Thus, pre-established polar nature of SBN supports active infrared absorption in SSBLN *nano* materials. In principle, such an absorption occurs due to change in the bond's dipole moment and due to which the typical bond absorbs infrared energy at certain frequency corresponding to its natural frequency of the vibration. Fig.3 shows the IR spectra of tin-doped SSBLN ceramics recorded in the wave number range from 500 to 4000 cm$^{-1}$. The spectrums for all samples are identical in nature suggesting that a) the expected double occupancy of $Sn^{2+}$ at sites of $Sr^{2+}$ in the SSBLN system is successful and b) it preserves the parent orthorhombic nature of the crystal structure. The position of IR-bands in low wavenumber regime (below 1000cm$^{-1}$) is an indicative of vibrating metal oxygen bonds, therefore, absorption bands close to 600 cm$^{-1}$ correspond to the *anti-symmetric* stretching vibrations of Sn-O-Sn bonds [15]. Overall, the infrared absorption bands observed in the range of 525-600 cm$^{-1}$ are due to resultant vibrations emerging from bending and stretching of Bi-O, Sr-O and Bi-O-Sr bonds. The absorption peak or transmittance dip at 795.20cm$^{-1}$ arises due to La-O bonds along with motion of oxygen sub lattice in polycrystalline SSBLN ceramics[16], besides a peak at 1033.86cm$^{-1}$ representing strong C-O stretching [17]. High transmittance in high wave number regime is a unique feature of current tin-doped SSBLN system compared to earlier reported SBT. This indicates about high strength of *bridged* TO (transverse optical)-LO (longitudinal optical) modes at high frequencies.

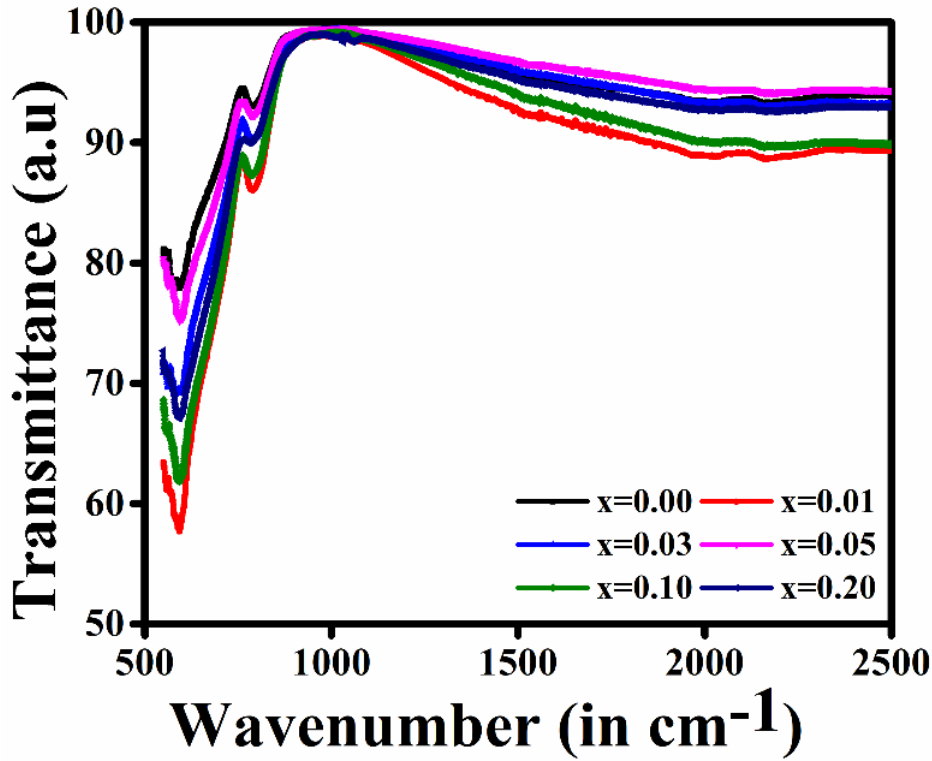

**Fig. 3** FTIR spectra for $Sr_{1-x}Sn_xBi_{1.95}La_{0.05}Nb_2O_9$ ceramics.

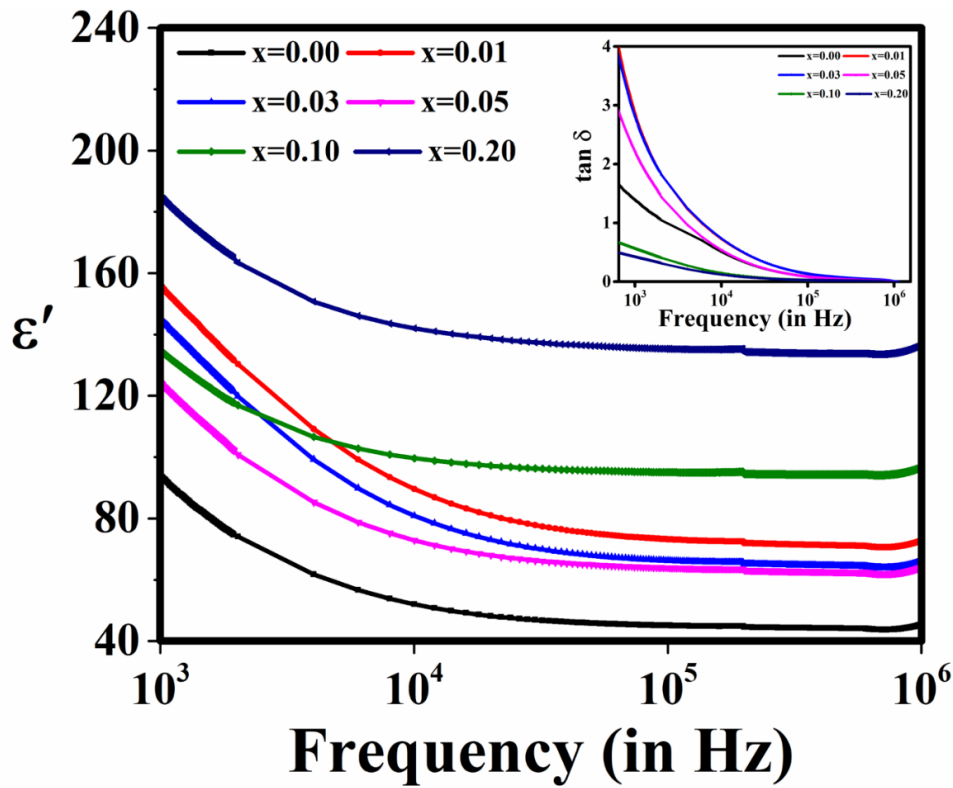

**Fig. 4 Dielectric dispersion of tin-doped SSBLN compositions.**

Tin as dopant is observed to increase the dielectric constant ($\varepsilon'$) of SSBLN compositions in a systematic manner for doping concentrations multiple of 5% (5, 10, 20). However, there was no analogy observed for 1% and 3% doped samples. The reason for irregular variation of dielectric constant/loss values for these samples could be the minimum unit cell change on doping [18-19]. Overall increase in $\varepsilon'$- values for all tin-doped samples compared to the undoped one is due to the presence of more polar covalent tin at the site of ionic-bonded strontium. This can also be seen in all dispersion curves of Fig.4 where orientation polarization plateau [20] possess higher $\varepsilon'$- values apart from regular typical increase in $\varepsilon'$-values on doping. A decrease in $\varepsilon'$-values with increase in frequency can be elucidated by the Koop's phenomenological theory based on the Maxwell-Wagner model considering inhomogeneous double layer dielectric structure [21]. The capacitive grain boundaries (responsive at low frequency electric fields) enclose conducting grains and produce shielding electric polarization known as a part space charge polarization other than interfacial one. Therefore, the values of $\varepsilon'$ are higher at lower frequencies due to thin capacitive layers those dissolve on increasing frequency mostly up to $10^5$ Hz in case of donor/acceptor-doped materials. Beyond frequencies $10^4$ Hz, all dispersion curves corresponding to orientation polarization follow an increasing trend on increasing tin-content for concentration multiple of 5 % (except the ones for 1 and 3 % tin). Nearly similar low frequency slopes of these dispersion curves for all samples provide an indication that substitution of $Sn^{2+}$ onto $Sr^{2+}$ sites does not create excess space charge in the SSBLN system compared to already produced during single stage sintering due to controlled bismuth oxide loss. This is possible only when introduced $Sn^{2+}$ ions a) successfully occupy $Sr^{2+}$ sites and b) maintain charge neutrality by fully engaging twelve-coordination number. The inset of Fig.4 shows the variation of loss tangent for all investigated tin-doped SSBLN materials. The loss values are higher for undoped SSBLN, x=0.01, 0.03 and

0.05 compared to samples with x=0.1 and 0.2. This can argued due to the presence of space charge in typical samples as explained earlier in this section. The magnitude of loss tangent values in conventional ferroelectrics is an indicative of combined space charge polarization and domain wall relaxation. Thus decrease in the loss values for samples with x=0.1 and 0.2 confirms an increase in the mobility of domain walls [18].

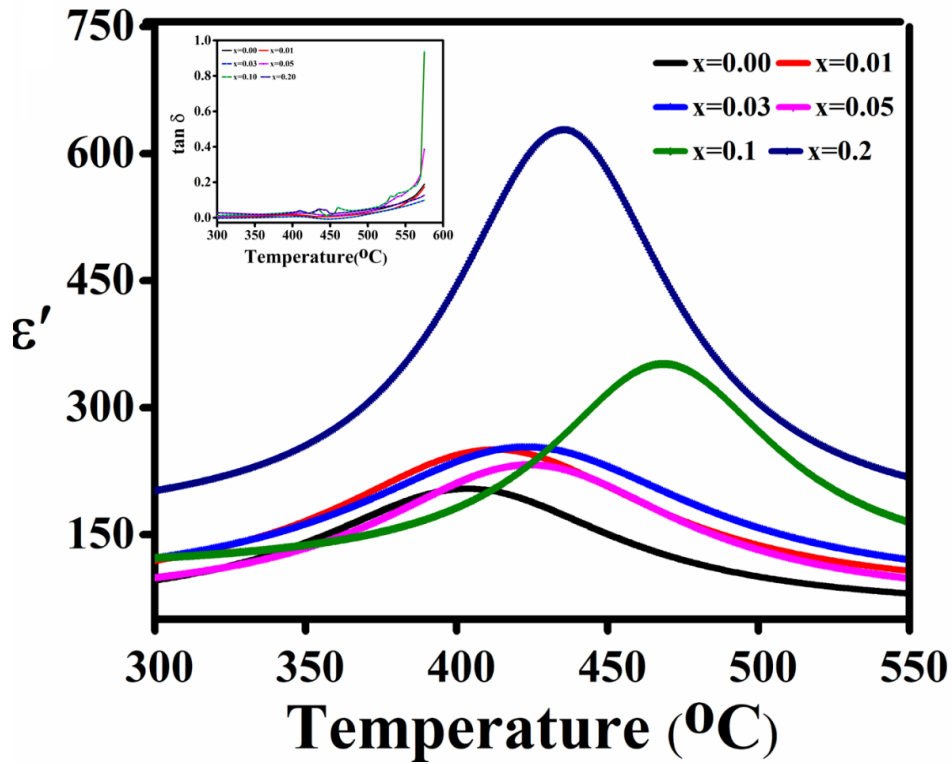

**Fig.5 Ferroelectric (FE)-paraelectric (PE) phase transition curves for $Sr_{1-x}Sn_xBi_{1.95}La_{0.5}Nb_2O_9$ samples (inset describes the variation of loss tangent).**

All tin-doped SSBLN compositions show typical single-phase ferroelectric (FE)-paraelectric (PE) phase transitions recorded at 1MHz frequency, Fig.5. A broad phase transition peak appears for all the samples that is an indicative of displace type transition occurring through polar *nano* regions (PNR). The phase transition Curie temperature is higher for all tin-doped samples compared to the undoped SSBLN and follow an increasing trend on increasing tin-content (*x*) [22]. There is slow variation in loss tangent values until 500ºC and rapid beyond this. This confirms the smooth domain wall relaxation and easy charge recovery between domains of these PNR based SSBLN materials and provides an indication of high electric fatigue resistance.[3]

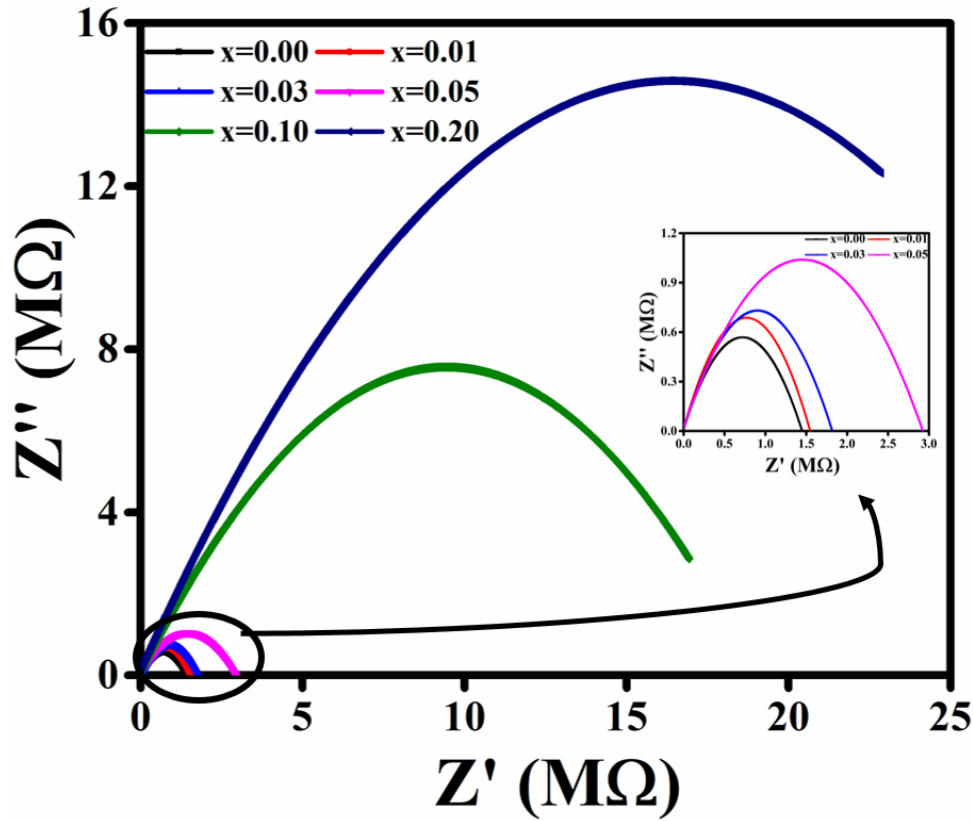

**Fig.6 Nyquist plots for investigated tin-doped SSBLN ceramics at room temperature**.

Fig.6 shows the complex impedance spectrum also known as Nyquist plots or Cole-Cole plots for tin-doped SSBLN ceramics at room temperature in the frequency range of 20Hz -2MHz. All plots possess single semicircular arc whose radii gradually increases on increasing tin content (*x*) until 5 atomic % and substantially increases thereafter for 10 and 20 atomic%. The intercepts of all these arcs on real impedance axis yield bulk resistance of typical tin-doped compositions. These single arcs represent single conduction mechanism in the material arising from grains rather than grain boundary and grain-electrode interface. These semicircular arcs shift towards high impedance values with increase in tin content ($x$) that is in accordance with discussions on Fig.5. This is interesting to notice that tin doping reinforces insulating covalent character rather than metallic one in SSBLN compositions. Nyquist plots as shown in Fig.6 have depression degree nearly on or slightly above the abscissa axis indicating non-ideal Debye formalism expected in most heterogeneous dielectrics [15,23-25]. In current SSBLN materials, atomic stoichiometric homogeneity is comparatively higher due to lanthanum controlled bismuth oxide evaporation than earlier reported SBN compositions prepared using other solid-state routes [26].

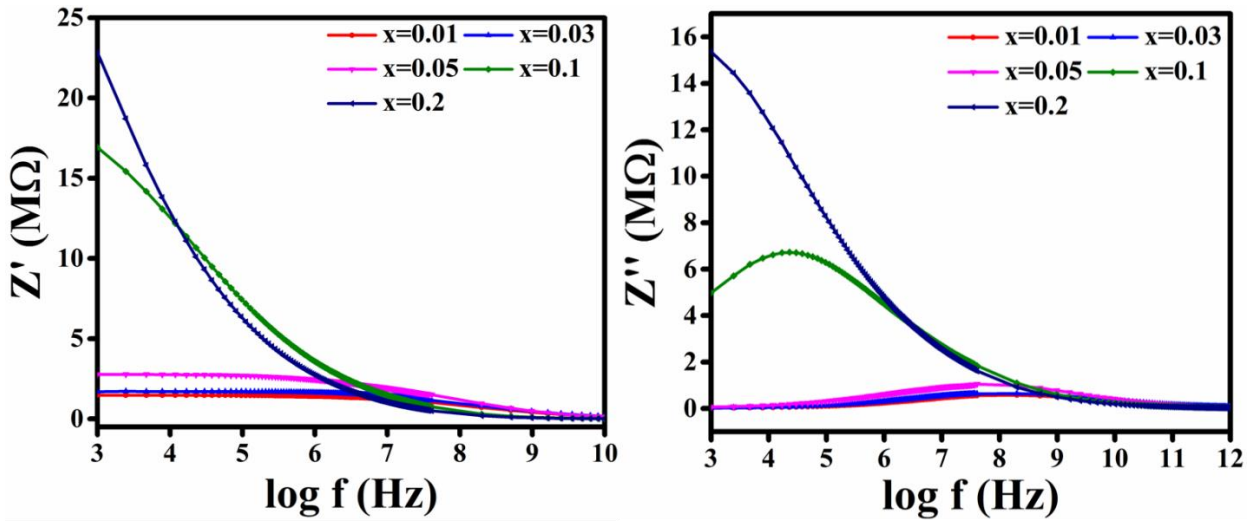

**Fig. 7 (a) Variation of real part of impedance (Z') with frequency (b) Variation of imaginary part of impedance (Z'') with frequency.**

Fig.7 shows typical (*a*) real and (*b*) imaginary impedance versus frequency behavior of all tin-doped SSBLN compositions. The magnitude of Z' decreases negligibly for tin-doped samples until 5% compared to rapid decrease in samples with 10 and 20% tin before finally merging at high frequencies. It can be concluded that real impedance Z' is independent of frequency and doping content (x) beyond 1kHz. The higher Z' values at low frequencies in samples with 10 and 20% tin is because of low mass space charge from PNR dominant microstructure of 10 and 20% tin doped SSBLN compositions [27-29], also seen in Fig.4. Fig.7(b) shows a very interesting variation of Z" with frequency for different doping concentrations. In general for all samples, complex impedance first increases with frequency, attains a maximum and decreases thereafter. This is a typical Non-Debye behavior of an ionic conductor and arises due to the presence of space charge in the system. The space charge saturates at typical frequency known as relaxation frequency. The relaxation frequency shifts towards lower frequencies on increasing tin content ($x$) further confirming presence of low mass negative space charge present is these typical 10 and 20% tin-doped SSBLN compositions. This is also an indicative of lower eddy current losses in higher tin-doped SSBLN samples. The presence of broad relaxation peak in low tin-doped ($\leq$ 5%) SSBLN samples impedance response is an indicative of strong *multiple* high frequency relaxations. Thus, low tin doping in SSBLN compositions can be extremely useful for high frequency broadband filter circuit applications.

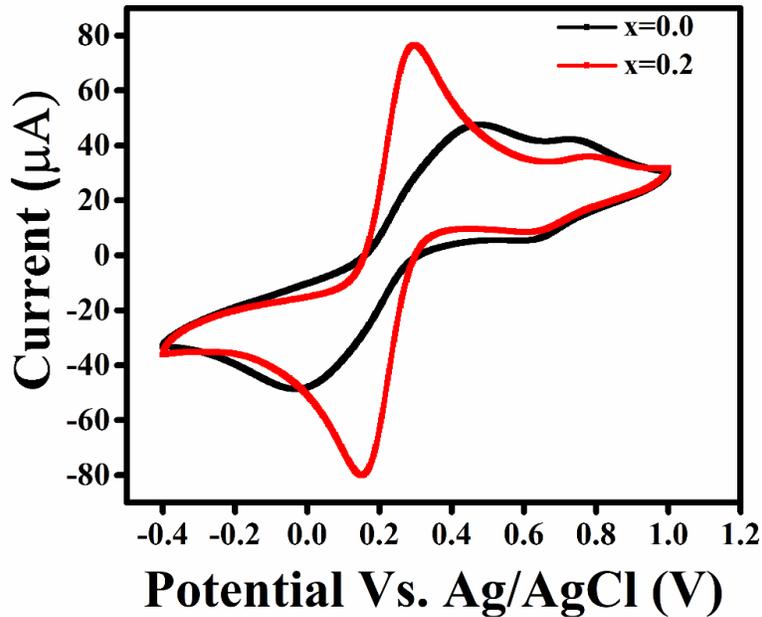

**Fig.8 Redox activity in typical undoped SBLN and 20% tin-doped SSBLN.**

Typical SSBLN compositions, undoped (SSBLN$_0$) and 20% tin-doped SSBLN (SSBLN$_{20}$) are modified using glassy carbon electrode (GCE) by dissolving 5 $m$M Fe(CN)$_6^{3-/4-}$ redox couple in 0.1 M KCl. A low peak potential difference ($\Delta E_p$= 0.1465 V) and high peak current (I=76µA) is observed in case of GCE/SSBLN$_{20}$ compared to GCE/SSBLN$_0$ ($\Delta E_p$= 0.5095 V and I=48µA), Fig.8. This is because of, *a*) large number of defect centers contributing as electron sources in undoped SSBLN, *b*) depleted defect sources as tin occupies Sr$^{2+}$ sites and imparts higher energy for holding covalent bonds with nearest neighbours (as a result redox peaks become sharper) and *c*) smooth electron transfer in GCE/SSBLN$_{20,}$ may be due to presence of double Sn$^{2+}$ on Sr$^{2+}$ site. The p-orbital electrons of Sn$^{2+}$ in the SSBLN system facilitate the quick electron transfer at electrode-electrolyte interface. This improved C-V behaviour of GCE/SSBLN$_{20}$ sample compared to undoped SSBLN is also an indicative of increased electrical conductivity of tin-doped samples besides having limited redox sources [30-31].

Due to opaque nature of the SBN ceramics, the diffuse absorption spectra is used to observe the change in the optical energy band gap of SSBLN ceramics in an integrating sphere mode of UV-vis spectrophotometer, Fig.9. The spectra contain significant optical band gap transition for the visible light absorption [32]. The *Kubelka-Munk* (K-M) function is used for computing band gap values for all samples: $\alpha h\nu = A(h\alpha - E)1/n$, where $\alpha$, $h$, $\nu$, $E$ and $A$ are the absorption coefficient, Planck's constant, frequency of light, band gap energy and a constant respectively ($n$-determines the transition feature in the semiconductor materials) [33]. The direct band gap is calculated by plotting $(\alpha h\nu)^2$ versus $h\nu$ as shown in inset of Fig.9 [34]. The values of regularly decreasing optical energy band gap with increase in tin content ($x$) are listed in Table-2. This confirms the *5p*-orbital tin contributing for lowering the conduction band edge and transforms dielectric SSBLN system into potential photovoltaic semiconducting SSBLN system.

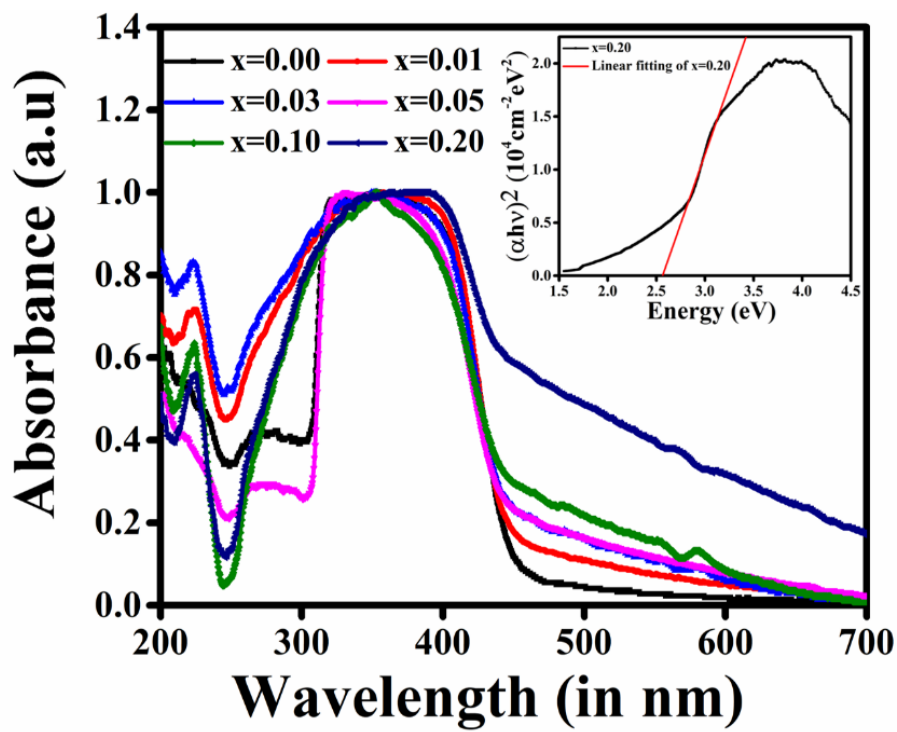

**Fig.9** UV-vis diffuse absorption spectra of SSBLN *nano* materials (inset depicts typical band gap calculation using Tauc method for 20% tin-doped SSBLN sample).

**Table-2** Variation of optical energy band gap of SSBLN ceramics with tin concentration

| Concentration of tin in SSBLN (in %) | Band gap (in eV) |
|---|---|
| 0 | 2.8119 |
| 1 | 2.8013 |
| 3 | 2.7265 |
| 5 | 2.6563 |
| 10 | 2.6374 |
| 20 | 2.5519 |

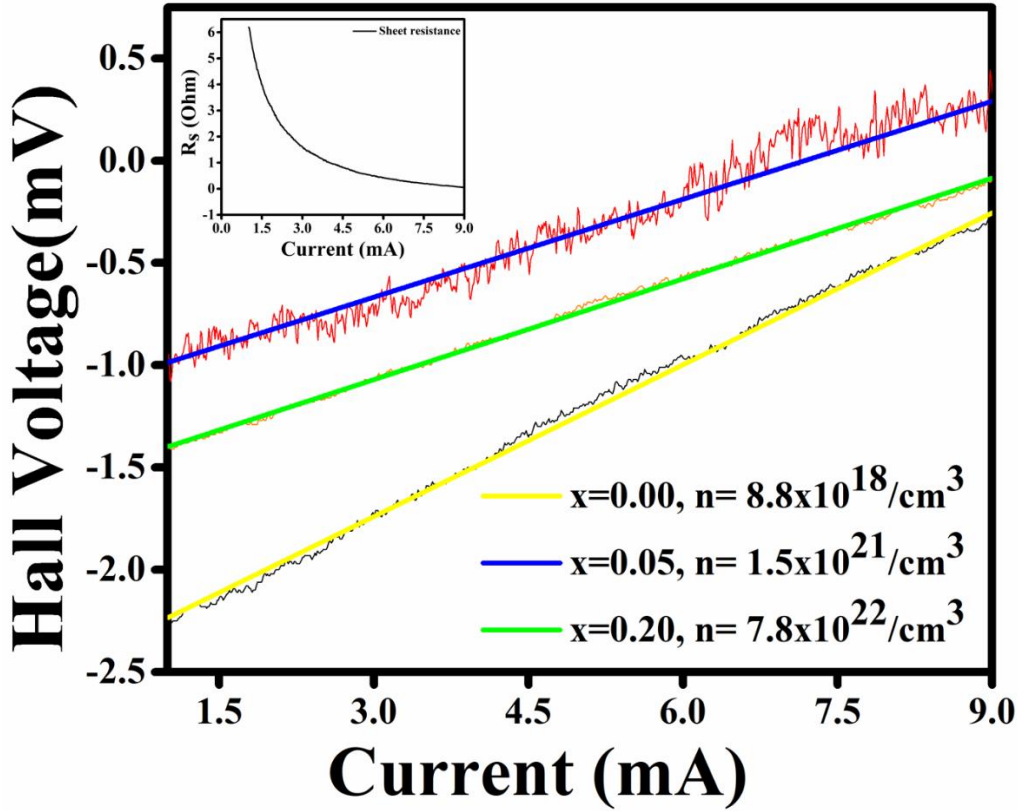

**Fig.10 Hall characteristics of a few SSBLN compositions at room temperature.**

Regularly decreasing optical energy band gap provides motivation of further study for perovskite photovoltaic applications using current SSBLN compositions. Recently many researchers[35-36] have highlighted the use of perovskite oxides like $TiO_2$ and $NiO_x$ for electron and hole transport layers (ETL-HTL). All reports indicate about importance of transport properties of semiconducting wafers under investigation and van der Pauw (vdP) method for this purpose. Therefore, room temperature Hall effect is investigated in a few randomly chosen SSBLN thin (0.5mm) pellets using van der Pauw (vdP) setup, Fig.10. The measurments are recorded at fixed magnetic field of 500 Gauss by decreasing current across two diagonally arranged perimeter contacts from 10 $m$A to 0.5$m$A. The other two diagonal arranged contacts are used for recording Hall voltage. First the Hall cofficient and later carrier mobility is calculated using composition dependent variation of

Hall voltage, Table-3. The sheet resistance is a critical parameter determining semiconducting wafer resistance in most photovoltaic materials. The sheet resistance of present SSBLN compositions changes in known conventional form for semiconducting/insulating wafers and is illustrated as an inset of Fig.10. The curves for initial tin doping up to 5% show an upward shift towards *positive* Hall voltage values indicating rise in *hole* concentration. However, there is a downward shift in Hall voltage data curves for higher tin concentrations 10 and 20 % indicating suppressed *hole* concentration. High mobility is recorded in high current range due to presence of high *hole* concentration [37].

**Table-3 Variation of resistivity and carrier mobility of SSBLN perovskites with tin content**

| Conc. of tin (in %) | Sheet resistance $R_s$ ($\Omega$.) | | Resistivity ($\Omega$-m) | Mobility (cm$^2$/V-s) | |
|---|---|---|---|---|---|
| | Below 3mA | Above 5mA | Below 3 mA | Below 3mA | Above 5mA |
| 0 | 4.54 | 0.43 | 3.18 | 1.09 | 11.52 |
| 5 | 1.86 | 0.13 | 1.30 | 1.72 | 24.40 |
| 20 | 2.80 | 0.23 | 1.96 | 1.17 | 14.02 |

## 4. Conclusions

Thorough ball milling along with unique microwave sintering yields *nano* particle driven grains in $Sr_{1-x}Sn_xBi_{1.95}La_{0.05}Nb_2O_9$ (SSBLN) materials. The change in tin concentration (*x*) is critical in determining various electrical and optical responses of investigated SSBLN compositions. The XRD patterns in correlation with HRTEM-EDS conclude the formation of unchanged orthorhombic phase with increase in unit cell volume indicating large *c*-axis displacement due to tin occupying $Sr^{2+}$ and $Bi^{3+}$ sites and double tin occupancy after $Bi^{3+}$ site saturation. Unchanged FTIR peak positions support the argument of unaltered orthorhombic perovskite phase with increased optical absorption and consistent metal-oxygen vibrations. Dielectric polarity was observed to increase significantly on tin doping due to tin being more polar covalent than strontium. Impedance spectroscopic investigations reveal single dominant grain contribution in conduction mechanism at the cost of dissolved grain boundary or electrode interfacing effects. This can be due to easy recovery of charge carriers between different grains and interfaces. Cyclic voltammetry measurements supplement the argument of absent grain boundary capacitive layers in tin-doped samples as appeared in case of undoped SSBLN in the form of multiple redox regimes. Reducing optical energy band gap values for investigated SSBLN system along with increasing Hall mobilities provided a strong indication for further utilizing current SSBLN composition for hole transport layer (HTL) *as well as* electron transport layer (ETL) in perovskite solar cells.

## Acknowledgements

The authors gratefully acknowledge the contributions of Dr.S.Amirthpandian (IGCAR-Kalapakkam) for providing HRTEM-EDS data. Authors also deeply acknowledge the contribution of Dr. S.S.Roy (SNU) for providing us cyclic voltammetry data and helping to prepare result analysis. One of the authors, Anurag Pritam, also acknowledges the gratitude towards Shiv Nadar Foundation for providing research fellowship.

**Electronic Supplementary File for the help of Reviewers**

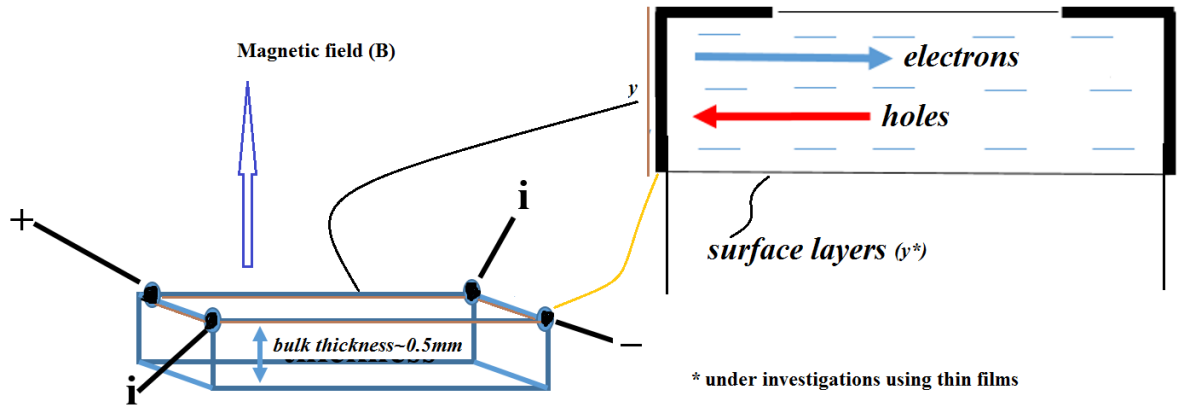